\documentclass[superscriptaddress,aps,prl,twocolumn]{revtex4}
\usepackage{amssymb}
\usepackage{graphicx}
\usepackage{natbib}
\usepackage{epstopdf}
\usepackage{dcolumn}
\usepackage{bm}
\usepackage{makeidx}
\usepackage{enumerate}
\begin{document}

\title{DNA Breathing Dynamics in the Presence of a Terahertz Field}
\author{B. S. Alexandrov}
\affiliation{Theoretical Division and Center for Nonlinear Studies, Los Alamos National Laboratory, Los Alamos, New Mexico 87545}
\author{V. Gelev}
\affiliation{Harvard Medical School, Boston, Massachusetts 02215}
\author{A. R. Bishop}
\affiliation{Theoretical Division and Center for Nonlinear Studies, Los Alamos National Laboratory, Los Alamos, New Mexico 87545}
\author{A. Usheva}
\affiliation{Harvard Medical School, Boston, Massachusetts 02215}
\author{K. \O. Rasmussen}
\affiliation{Theoretical Division and Center for Nonlinear Studies, Los Alamos National Laboratory, Los Alamos, New Mexico 87545}
\date{\today}

\begin{abstract}
We consider the influence of a terahertz field on the breathing dynamics of double-stranded DNA. We model the spontaneous formation of spatially 
localized openings of a damped and driven DNA chain, and find that linear instabilities lead to dynamic dimerization, while true local strand separations
require a threshold amplitude mechanism. Based on our results we argue that a specific terahertz radiation exposure may significantly 
affect the natural dynamics of DNA, and thereby influence intricate molecular processes involved in gene expression and DNA replication.
\end{abstract}
\maketitle
Spectroscopic techniques in the terahertz range are currently emerging as new tools for the investigation of biological macromolecules \cite{Markelz}. In spite of 
the experimental difficulties (caused by the water's giant Debye dipole moment, which leads to a substantial dielectric relaxational 
loss in the THz range) a notable amount of research effort is being devoted to the development of sophisticated THz bio-imaging \cite{ Nature Photonics}. Nevertheless,
very little is known about THz-radiation's influence on biological systems, and the mechanisms that govern this influence. The possibility that low frequency electromagnetic 
radiation may affect genetic material, enzymatic reactions, etc. was introduced long ago \cite{Frohlich}, and since then has been a subject of constant debate. The energy of 
such radiation is too low to directly disrupt any chemical bonds or cause electronic transitions. Only a resonance-type interaction might lead to an appreciable, biological effect. 
In biomolecules such interactions are possible through the ubiquitous hydrogen bonds that have energies in the THz range. Numerous {\it in vivo} and {\it in vitro} experiments
have been conducted to clarify low frequency radiation's ability to cause biological effects, such as chromosomal aberration, genetic damage etc. 
The experimental studies have been conducted under a variety of conditions, but mostly at frequencies below 0.01 THz, power below 1 $mW/cm^2$, and short exposure times. The 
data collected in these conditions led to mixed conclusions: some studies reported significant genetic damages while others, although similar, showed none
\cite{Mutation Research}. The major international research project, "THz-bridge" \cite{THz-Bridge}, which was specifically concerned with THz radiation 
genotoxicity concluded that: {\it under some specific conditions of exposure, change in membrane permeability of liposomes was detected and an induction of genotoxicity was observed to occur in
lymphocytes}. Hence,  this project confirmed the existence of THz genotoxicity, but it remains unclear under which specific conditions such effects 
occur. 

Recent measurements confirm that only extended (6 hours) exposure to a weak THz field can cause genomic instability in human lymphocytes \cite{Korenstein}. 
Independently, it was reported that neurons exposed {\it in vitro} to powerful THz radiation (over 30 $mW/cm^2$ ) cause infringement of the morphology of the 
cellular membranes and intracellular structures \cite{Russ1}. The same work also showed that at decreasing power and/or at different frequencies the 
morphological changes do not occur. Recently, it was further pointed out that exposure to a low level THz radiation can interfere with the 
protein-recognition processes \cite{Korenstein II}. It was also shown that exposure of mice to 3.6 THz high power (15 $mW/cm^2$) radiation, for 
30 min caused behavioral changes \cite{Russ2}, while under short (5 min) exposure the changes could not be detected. 
\begin{figure}[h]
\includegraphics[width=\columnwidth]{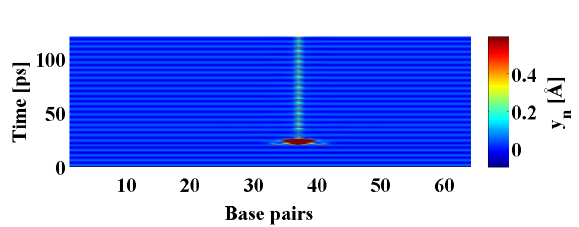}
\caption{Spontaneous formation of a localized permanent opening in dsDNA in the presence of a THz field with amplitude $A=144$ pN, and $\Omega=2$ and a spatial  fluctuation with high amplitude (see text).} 
\label{fig0}
\end{figure}
Thus the available experimental data strongly suggest that THz-radiation can affect biological function, {\it but  only under specific conditions}, 
viz. high power, or/and extended exposure, or/and specific THz frequency. 

The present work is devoted to developing a qualitative description of the existence of such THz effects. We argue that the appropriate 
conditions correspond to irradiation parameters that assure the existence of linear and nonlinear resonances between DNA 
conformational dynamics and THz field Fig.\ref{fig0}. 

We will base our analysis on the Peyrard-Bishop-Dauxois (PBD) model of dsDNA \cite{2}, which is arguably the most successful available model for describing the local DNA pairing/unpairing (breathing) dynamics \cite{6}. Of particular importance for the THz effect is our previous demonstration of strong correlations between regulatory activity, such as protein-DNA  binding and 
transcription, and the {\it equilibrium} propensity of dsDNA for local strand separation \cite{-8,3,4,5}. With this background it is natural to consider the influence of THz radiation on 
dsDNA dynamics within the PBD modeling framework. One complication is that the specific physical nature of the interactions between DNA and the THz electromagnetic field is not known in 
detail. However, given the sensitivity of THz radiation to the strand pairing state of DNA \cite{15},  we will here simply augment the PBD to include a drive in the THz frequency range, 
without specifying the precise nature of the underlying physical coupling. Since the wavelength of the THz field is larger than 100 $\mu m$ and the characteristic size the investigated
dsDNA sequences is less than 20 $nm$ it is reasonable to consider an uniform THz field. In this fashion the breathing dynamics (represented by the normalized separation $y_n$ of complementary 
bases) of the $n$'{th} base pair in the presence of a monochromatic spatially homogeneous external field is described by
\begin{eqnarray} \label{eqn2}
m \ddot y_n = &-& U^\prime (y_n) - W^\prime (y_{n+1},y_{n})
- W^\prime (y_{n},y_{n-1})\nonumber \\ &- &m\gamma \dot y_n +A\cos{(\Omega t)},
\end{eqnarray}
where, the Morse potential $U (y_n)=D_n(\exp {(-a_ny_n)} -1)^2$ represents the hydrogen bonding of the complementary bases. The parameters $D_n$ and $a_n$ depend on the type of the base pair (A-T or G-C).  
Similarly, $W (y_n, y_{n-1})=\frac{1}{2}\chi[y_n,y_{n-1}](y_n-y_{n-1} )^2$ with $\chi[y_n,y_{n-1}]= k(1+\rho \exp {(-\beta(y_n - y_{n-1})) })$ represents the stacking energy between consecutive base pairs.  The term $m\gamma \dot y_n$ is the drag caused by the solvent while $A\cos{(\Omega t)}$ is the (THz) drive.  For simplicity, we consider here a homogeneous poly(A) \cite{note_on_parameters} DNA molecule with 64 base pairs. Our numerical simulations of this set of nonlinear coupled equations (using periodic boundary conditions) showed that the primary response of the system is spatially homogeneous (Fig.\ref{fig0} the first 40 $ps$).  However, we observed that a presence of fluctuations may lead to the creation of persistent spatially localized openings of the double-stranded molecule, Fig.\ref{fig0}. Because,
such openings (bubbles) are known to functionally affect dsDNA we focus here on characterizing the conditions under which these states appear. 

\begin{figure}[h]
 \includegraphics[width=\columnwidth]{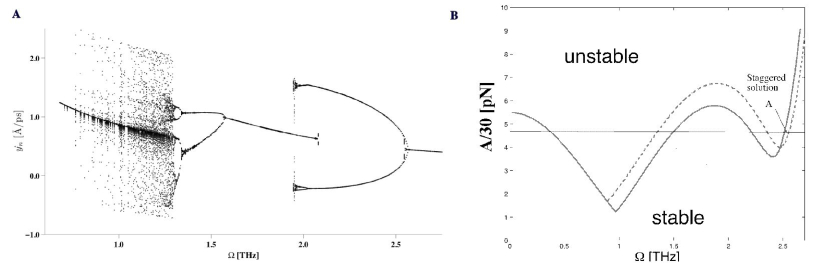}%
 \caption{A) Bifurcation diagram for a poly(A) molecule in an external THz field with amplitude $A=144$ pN. B) The instability curves for wave vector $k= 0$ (solid curve) and $k=\pi$ (dashed curve) are plotted. On the vertical axis is the amplitude of the external field, and on the horizontal axis is the frequency $\Omega$ of the THz  field. }
\label{fig1}
\end{figure}

Since, the primary response of of the driven system is spatially uniform it can be understood in terms of a single classical damped and driven Morse oscillator. Such systems have been studied quite extensively in 
many other contexts \cite{7}, and are known to display rich dynamical behavior including limit cycles, period-doubling bifurcations etc. An illustration of this is shown in  Fig.\ref{fig1}, A, for our 
system parameters, in terms of the Poincar{\'e} section $\dot y_n(t=mT)$, where $T=2\pi/\Omega$, versus the driving frequency $\Omega$ for several integer values of $m$. Feigenbaum's period-doubling 
route to chaos is clearly seen in this figure, which demonstrates that this system's main instability mechanism is through period-doubling.
As described earlier, our interest is to investigate whether the presences of a THz field can lead to the creation of spatially localized unbinding of the DNA double-strand. Such states are most easily 
created in connection with the period-doubling events, as a result of a spatially inhomogeneous perturbation. In order to investigate this phenomenon we adopt a method developed in 
Ref. \cite{9}. In this work the linear stability analysis (based on the Floquet theorem) of a single Morse oscillator was performed, and it was shown that in a rotating-wave approximation \cite{Bessel} the dynamics ensuing from a given perturbation is governed by the extended Hill equation. The stability analysis of the Hill equation has been developed by Ince  \cite{11} based on his extension of the Whittaker's method of solving Mathieu's equation \cite{20}. 
Since, our 
system is spatially extended, a perturbation may be spatially inhomogeneous. In this case the response of the system
is characterized by $y_n(t)=y^0(t)+z_n(t)$, where $y^0(t)$ is the spatially homogeneous and temporally periodic solution described above 
(see Fig. \ref{fig1}), A, and $z_n(t)$ represents the perturbations in whose 
dynamics we are interested. In the most general form the perturbation is given in terms of Fourier modes: $z_n(t) =\sum_k \exp [ ikn] \xi_k(t)$ with wavenumber, $k$, and amplitude $\xi_k(t)$. 
As with the single oscillator, the dynamics of the Fourier amplitudes $\xi_k(t)$  of the perturbation are governed by set of uncoupled extended Hill equations. However, the coefficients of these equations depend on the 
wave-number $k$. The Ince stability analysis allows us to analytically determine \cite{note1} the Floquet exponents (i.e. the stability criteria) as functions of the spatial profile as given by $k$. For more technical 
details on the method see Ref. \cite{12}. The 
result of our analysis are the stability curves dividing  the parameter space $(A/m, \Omega, k)$ into stability and instability zones. Figure \ref{fig1}, panel B shows the stability boundaries of a spatially homogeneous perturbation 
($k=0$, solid curve) and of a staggered ($k=\pi$, dashed curve) perturbation. Perturbations represented by other wave-numbers have stability boundaries in-between those 
shown in Fig. \ref{fig1}, B. Below these curves the spatially homogeneous states perform stable temporal oscillations at the frequency of the external drive and its higher 
harmonics, while above the curve the 
perturbation leads to period-doubled oscillations. It is clearly seen that for driving frequencies below $\Omega \sim  2.5$ THz spatially homogeneous ($k=0$) perturbations lead to the strongest instability (occurring 
at the lowest value of the driving amplitude). However, above $\Omega \sim  2.5$ THz the staggered ($k=\pi$) perturbation becomes the primary instability mode. In the region 
above the stability boundary the period-doubled state will therefore tend to adopt the spatial profile of the most unstable Fourier-mode (homogeneous or staggered). 
The presence and location of the 
\begin{figure}
\includegraphics[width=\columnwidth]{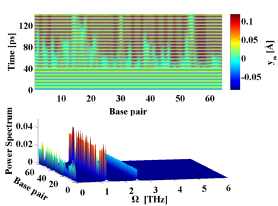}%
\caption{The $k=\pi$ spatial pattern, at $\Omega =2.52$ THz, is presented. The upper panel shows evolution of the system. The lower panel shows the power spectrum of the breather motion.}
\label{fig3}
\end{figure}
\begin{figure}
\includegraphics[width=\columnwidth]{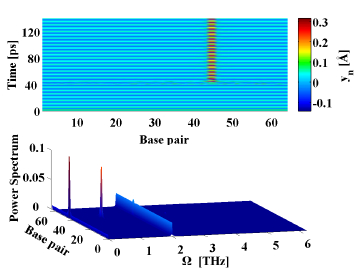}
\caption{Breather formation at $\Omega =2.00$ THz. The upper panel shows the evolution of the breather. The lower panel shows the power spectrum of the breather motion.}
\label{fig4}
\end{figure}
obtained instability curves (Fig. \ref{fig1}, B) were verified by exploring  the parameter-space $(A/m, \Omega,k)$ by direct simulations of Eq. (1). 
The dynamics of small spatially random (uniformly distributed in the interval $[-\epsilon, \epsilon]$ and thus containing all wave-numbers) 
perturbation $z_n$ introduced on top of the spatially homogeneous and temporally periodic solution $y^0(t)$ was followed numerically.
As this perturbation injects energy into all wave-numbers, the ensuing dynamics will emphasize the mode with the strongest instability. 
For point A (Fig. \ref{fig1}, B), the results are shown in Figure \ref{fig3}. The upper panel shows that after the introduction of the perturbation (at 40 ps)
a spatially staggered structure forms. The power spectrum of this dynamics (lower panel in Fig. \ref{fig3}), shows that the frequency of the vibrations is mainly $\Omega/2$, i.e. a 
period-doubling transition has occurred as expected from the above analysis. The power spectrum (lower panel) shows that the signature of the drive persists 
at $\Omega$, in addition to a component at zero frequency, which results from the asymmetry of the base-pair potential $U(y_n)$.
A similar scenario is observed for driving frequencies below $\Omega \sim 2.5$ THz, but in these cases the solutions remain spatially homogeneous after the 
introduction of the perturbation.
The fact that the instability curves for $0< k <\pi $ lie between the two curves in Fig. \ref{fig1}, B shows that it is, in general, impossible for any spatially localized feature with a finite ($ 2~\mbox{base pairs}  < \lambda < \infty$) 
length scale $\lambda$ to emerge as a result of a linear instability. This is in contrast with many other nonlinear condensed matter models, where spatially localized and temporally periodic 
modes (intrinsic localized modes or breathers) commonly arise through linear modulational instabilities. For our model of dsDNA a perturbation of finite 
amplitude $\epsilon$ is required to create localized unbindings of the DNA double strand through nonlinear mechanisms. An illustration of this is given in Fig. \ref{fig4} for $\Omega = 2.0$ THz, where the dynamics caused 
by a random perturbation with $\epsilon > 0.14$ is shown to result in a localized opening of the DNA double strand. 
Again the power spectrum, given in the lower panel of Fig. \ref{fig4} shows, as expected, that the localized state vibrates at the frequency $\Omega/2$ in contrast to the uniform background, which remains synchronized to the external drive at $\Omega$. 
A more detailed study of the nonlinear mechanisms underlying the creation of such localized unbinding states reveals that a localized injection of a certain amount of energy is required. Due to the asymmetry of the base pair potential, an effective way to achieve this is 
a localized {\em compression} of the double strand. Using this insight we found the perturbation $z_n=0.42\cos{(\frac{\pi}{4}(n-n_0))}$, for $-4 \leq n-n_o \leq 4$ ($z_n=0$ otherwise) to be an effective perturbation for the 
creation of a localized unbinding state at $n_0$, Fig.\ref{fig0}. The phase portrait (not shown) of the base pair $n_0$, at the transient time period just before the breather is established, reveals the breather formation mechanism. First, the locally injected energy from the {\em compression} must be sufficient for a few consecutive base pairs to undergo a temporary local melting transition, i.e. reach 
the plateau of the Morse potential. Subsequently, the stacking interactions and friction dissipate the energy to gradually reach a steady state. At given driving amplitude and frequency, 
the dsDNA molecule has two 
steady conformational states with different average energies and amplitudes. Only in these states is the externally pumped energy exactly compensated by the energy dissipated by the friction. The first state is vibrating predominantly at the 
driving frequency $\Omega$, while the second mainly oscillates at $\Omega/2$. Hence, the breather formation proceeds in three stages: 1) all the base pairs are in the first state then; 2) a few of them undergo a temporary melting  transition as a result of the perturbational energy injection. Finally, 3) the perturbed base pairs reach the second steady conformational state, after dissipating their additional energy. 

In summary, we have found that a THz field may cause dynamical separations of the DNA double-strand.
In the presence of weak perturbations e.g. thermal fluctuations, small amplitude response occurs (at half the driving frequency) either in a spatially uniform manner or, at higher-frequencies, in an unusual spatially  dimerized form. In the latter case, neighboring base pairs oscillate in an out-of-phase fashion. However, large localized openings (bubbles) in the DNA double strand can only occur via a nonlinear mechanism  requiring a spatial perturbation above a certain amplitude threshold that is determined by the intensity and the frequency of the THz field. 

We previously showed experimentally that the introduction of an artificial {\it permanent bubble} in dsDNA, via a short mismatched segment, is in fact sufficient for 
transcription even in the absence of any auxiliary factors \cite{5}. Our present finding of large localized openings resulting from an external THz field therefore underscores the 
importance of including the interactions of genomic DNA with the surrounding environment in DNA models. The  amplitudes of the resulting openings observed in the presence of a THz 
field are significant compared to the dynamic signatures of protein binding sites and transcription start sites \cite{4}. This suggests that THz radiation may significantly interfere 
with the naturally occurring local strand separation dynamics of double-stranded DNA, and consequently, with DNA function. 

Based on the model results present here, we believe that the main effect of THz radiation is to resonantly influence the dynamical stability of the dsDNA system. Hence, 
our instability curves define the parameters space (i.e. the amplitude of the field (or the power), and frequency) in which
THz radiation can have an immediate effect. In contrast, nonlinear instability may occur at any point of the instability diagram but it requires fluctuations with significant amplitudes. 
In biological systems, such fluctuations are generated thermally. Hence, the occurrence of a fluctuation with sufficiently large amplitude is very rare, and therefore extended exposure is 
required for the THz effect to take place, via nonlinear instabilities, especially if the power is small. In this framework, it is than natural
that the character of THz genotoxic effects are probabilistic rather than deterministic.

We acknowledge Dr. Voulgarakis for initial discussions regarding the subject of this work.This research was carried out under the auspices of the U.S. Department of Energy at Los Alamos National Laboratory under Contract No. DE-AC52-06NA25396. and it was supported by the National Institutes of Health (R01 GM073911 to A.U.).



\begin{thebibliography}{12}
\bibitem{note_on_parameters} The specific parameter values are: $D_n$ =0.05 eV, $a_n$ = 4.2 $\AA^{-1}$, k=0.025 eV$\AA^{-2}$, $\beta$=0.35 $\AA^{-1}$, and $\rho$=2. The linear resonance frequency is $\omega=$1.22 THz. For the solvent drag we use $\gamma= 1$ps$^{-1}$, consistent with a typical relaxation time of water.
\bibitem{Bessel}In this specific rotating-wave approximation the Morse potential is expanded using the equality: $\exp{(r\cos(\phi))}=I_0(r)+2\sum^\infty _{n=1}{I_n(r)\cos(n \phi)}$ 
where $I_n(r)$ are modified Bessel functions of the first kind.
\bibitem{note1} For the analytical work we used a harmonic stacking potential, $\chi= const$.
\end{thebibliography}
\end{document}